\documentclass[natbib,epj]{svjour}

\usepackage{amsmath}
\usepackage{amssymb}
\usepackage{amsfonts}
\usepackage{graphicx}
\usepackage{dcolumn}
\usepackage{bm}
\usepackage{color}
\usepackage{multirow,longtable}
\usepackage{times}

\begin{document}

\title{Statistical properties of volatility return intervals of Chinese stocks}%

\author{Fei Ren\inst{1,2,}{\color{blue}{\thanks{e-mail:
fren@ecust.edu.cn}}} \and Liang Guo\inst{1} \and Wei-Xing
Zhou\inst{1,2,3,4,}{\color{blue}{\thanks{e-mail:
wxzhou@ecust.edu.cn}}}}

\institute{School of Business, East China University of Science and
Technology, Shanghai 200237, China \and Research Center for
Econophysics, East China University of Science and Technology,
Shanghai 200237, China \and School of Science, East China University
of Science and Technology, Shanghai 200237, China \and Research
Center of Systems Engineering, East China University of Science and
Technology, Shanghai 200237, China}

\titlerunning{Statistical properties in the volatility return intervals of Chinese stocks}


\date{Received \today  / Revised version: }

\abstract{The statistical properties of the return intervals
$\tau_q$ between successive 1-min volatilities of 30 liquid Chinese
stocks exceeding a certain threshold $q$ are carefully studied. The
Kolmogorov-Smirnov (KS) test shows that 12 stocks exhibit scaling
behaviors in the distributions of $\tau_q$ for different thresholds
$q$. Furthermore, the KS test and weighted KS test shows that the
scaled return interval distributions of 6 stocks (out of the 12
stocks) can be nicely fitted by a stretched exponential function
$f(\tau/\bar{\tau})\sim e^{- \alpha (\tau/\bar{\tau})^{\gamma}}$
with $\gamma\approx0.31$ under the significance level of 5\%, where
$\bar{\tau}$ is the mean return interval. The investigation of the
conditional probability distribution $P_q(\tau | \tau_0)$ and the
mean conditional return interval $\langle \tau| \tau_0 \rangle$
demonstrates the existence of short-term correlation between
successive return interval intervals. We further study the mean
return interval $\langle \tau| \tau_0 \rangle$ after a cluster of
$n$ intervals and the fluctuation $F(l)$ using detrended fluctuation
analysis and find that long-term memory also exists in the
volatility return intervals.
\PACS{
      {89.65.Gh}{Economics; econophysics, financial markets, business and management}   \and
      {89.75.Da}{Systems obeying scaling laws}   \and
      {05.45.Tp}{Time series analysis}
     }
}

\maketitle

\section{Introduction}
\label{intro}

In recent years many concepts and methods from statistical physics
have been applied to the study of financial markets
\cite{Mantegna-Stanley-2000}. The statistical analysis of the
waiting time between two successive events has drawn much attention.
Different definitions of {\em{event}} refer to a variety of
variables characterizing the properties of stock markets from
certain aspects. For example, the persistence probability concerns
the duration time that prices (or volatilities) always keep above or
below their initial values
\cite{Zheng-2002-MPLB,Ren-Zheng-2003-PLA,Ren-Zheng-Lin-Wen-Trimper-2005-PA},
the exit time raises a concept similar to the persistence
probability and focuses on the interval time between prices with
fluctuations larger than a certain threshold
\cite{Simonsen-Jensen-Johansen-2002-EPJB,Jensen-Johansen-Simonsen-2003-IJMPC,Jensen-Johansen-Simonsen-2003-PA,ZaluskaKotur-Karpio-Orlowski-2006-APPB,Karpio-ZaluskaKotur-Orlowski-2007-PA,Zhou-Yuan-2005-PA},
and the intertrade duration stands for the time interval between two
successive independent trades
\cite{Scalas-Gorenflo-Luckock-Mainardi-Mantelli-Raberto-2004-QF,Ivanov-Yuen-Podobnik-Lee-2004-PRE,Sazuka-2007-PA,Jiang-Chen-Zhou-2008-PA}.

Recently, extensive work has been done in the analysis of waiting
time between extreme events in natural records, such as floods,
temperatures and earthquakes
\cite{Bunde-Eichner-Havlin-Kantelhardt-2003-PA,Bunde-Eichner-Havlin-Kantelhardt-2004-PA,Bunde-Eichner-Kantelhardt-Havlin-2005-PRL,Saichev-Sornette-2006-PRL}.
The understanding of the statistical properties of these extreme
events is of great importance for the risk assessment of rare
events. By investigating the return intervals between successive
extreme events exceeding a certain threshold $q$, scaling behaviors
are revealed in the return interval distributions for numerous
complex systems. The dependence of the return interval distribution
on the correlation of the original records has been carefully
investigated
\cite{Pennetta-2006-EPJB,Olla-2007-PRE,Eichner-Kantelhardt-Bunde-Havlin-2006-PRE,Eichner-Kantelhardt-Bunde-Havlin-2007-PRE,Bogachev-Eichner-Bunde-2007-PRL}.
It shows that the scaling behavior of the return intervals might be
due to the long-term memory of the original records. It is well
known that long memory is an important statistical property
basically observed in stock volatilities
\cite{Mantegna-Stanley-2000}. Therefore, one can expect that the
scaling behavior might also appear in the volatility records of
stocks.

A first effort was conducted by Yamasaki {\em{et al.}}, who used the
daily data of US stocks to study the properties of the volatility
return intervals
\cite{Yamasaki-Muchnik-Havlin-Bunde-Stanley-2005-PNAS}. Suppose that
$Y(t)$ is the price at time $t$, and the volatility is simply
defined as the magnitude of logarithmic return $R(t,\delta
t)=|\ln(Y(t))-\ln(Y(t-\delta t))|$. They studied the return
intervals $\tau$ between successive volatilities above a certain
threshold $q$, and found that the distribution of volatility return
intervals $P_q(\tau)$ indeed obeys a scaling behavior. Wang {\em{et
al.}} further studied the statistical properties of the volatility
return intervals using the intraday data of US stocks
\cite{Wang-Yamasaki-Havlin-Stanley-2006-PRE,Wang-Weber-Yamasaki-Havlin-Stanley-2007-EPJB,VodenskaChitkushev-Wang-Weber-Yamasaki-Havlin-Stanley-2008-EPJB}.
They found that $P_q(\tau)$ exhibited a similar scaling behavior and
the distribution could be approximated by a stretched exponential
form similar to that found in climate records
\cite{Bunde-Eichner-Kantelhardt-Havlin-2005-PRL}. All these studies
show that the volatility return intervals have long memory. Similar
scaling behavior and long-term memory are observed in the daily and
1-min volatility return intervals of thousands Japanese stocks
\cite{Jung-Wang-Havlin-Kaizoji-Moon-Stanley-2008-EPJB}. Qiu {\em{et
al.}} studied the volatility return intervals using the
high-frequency intraday data of four liquid stocks traded in the
Chinese market, which is an important emerging market, and found the
return interval distribution also follows scaling behavior
\cite{Qiu-Guo-Chen-2008-PA}.

Contrary to the scaling behavior found previously, Wang {\em{et
al.}} analyzed the trade-by-trade data of $500$ stocks which compose
the S\&P 500 index and found that the cumulative distribution of
return intervals had systematic deviations from scaling and showed
multiscaling behaviors \cite{Wang-Yamasaki-Havlin-Stanley-2008-PRE}.
The analysis was based on the deduction that if the probability
distribution obeys scaling behavior its cumulative distribution
should also obeys scaling behavior. They further studied the $m$-th
moment of the scaled return intervals, and found the moment showed a
certain trend with the mean interval which supports the finding that
the return interval distribution exhibits a multiscaling behavior.
This remarkable multiscaling behavior calls for a more careful
scrutiny of the volatility return intervals by adopting solid
statistical methods rather than eyeballing. In addition, Lee {\em{et
al.}} investigated the return intervals of 1-min volatility data of
the Korean KOSPI index \cite{Lee-Lee-Rikvold-2006-JKPS}. They found
that the interval return distribution had a power-law tail and no
scaling was observed. However, it seems that they did not remove the
intraday pattern from the intraday volatility series, which weakens
their conclusion. Note that intraday pattern should be removed when
dealing with intraday data, which will have heavy influence on the
estimated return intervals for large thresholds.

In this work, we use a nice high-frequency database
\cite{Jiang-Guo-Zhou-2007-EPJB} to study the statistical properties
of the interval returns. The two samples Kolmogorov-Smirnov (KS)
test, which is a standard way to check whether these two samples are
drawn from a same distribution by comparing their cumulative
distributions, is used to examine the coincidence of the interval
distributions for different values of threshold $q$. Specifically,
we study in detail 30 most liquid stocks to gain better statistics.
Our results show that only $12$ individual stocks pass the KS test
and show scaling behaviors, and the remaining $18$ stocks show
nonscaling behaviors which seems consistent with that found by Wang
{\em{et al.}} \cite{Wang-Yamasaki-Havlin-Stanley-2008-PRE}.
Furthermore, a KS goodness of fit test is performed to study the
particular form of the scaling function. The memory effect is also
investigated in this paper.

The paper is organized as follows. In Section \ref{S1:Data}, we
explain the database analyzed and how the volatility return
intervals are calculated. Section \ref{S1:PDF} studies the scaling
and nonscaling behaviors of the volatility return interval
distributions using the KS tests. In Section \ref{S1:Memory}, we
further study the memory effect of volatility return intervals.
Section \ref{S1:concl} concludes.

\section{Preprosessing the data sets}
\label{S1:Data}

Our analysis is based on the high-frequency intraday data of $30$
most liquid stocks traded on the Shanghai Stock Exchange and the
Shenzhen Stock Exchange. These $30$ stocks are most actively traded
stocks representative in a variety of industry sectors, and thus
have the largest sizes among all the stocks. The basic information
about these stocks including the stock codes, company names and
industrial sectors they belonging to is shown in
Table~\ref{TB:two-sample-KS}. The dada record the trading prices
every six to eight seconds from January 2004 to June 2006. The
volatility is defined as the magnitude of logarithmic price return
between two consecutive minutes, that is
$R(t)=|\ln(Y(t))-\ln(Y(t-1))|$, where the price is the closest tick
to a minute mark. Thus the sampling time is one minute, and the
volatility data size is about 140,000 for each stock.

\begin{table*}[htb]
\caption{Kolmogorov-Smirnov test of the return interval
distributions for $q=2$ and $5$ by comparing the statistic $KS$ with
the critical value $CV$ at the 5\% significance level.}
\label{TB:two-sample-KS}
\centering
\begin{tabular}{cllccc}
\hline
Code & Stock name & Industrial sector & $KS$ & $CV$ & Scaling?\\
\hline
$000002$ & China Vanke Co., Ltd & Real estate & $0.0793$ & $0.0636$ & No\\
$000503$ & Searainbow Holding Co., Ltd & Conglomerates & $0.0485$ & $0.0469$ & No\\
$000625$ & Chongqing Changan Automobile Co., Ltd & Manufacturing & $0.0233$ & $0.0509$ &Yes\\
$000839$ & Citic Guoan Information Industry Co., Ltd & IT & $0.0705$ & $0.0490$ & No\\
$000858$ & Wuliangye Yibin Co., Ltd & Wine & $0.0803$ & $0.0500$&No\\
$000917$ & Hunan TV\&Broadcast Intermediary Co., Ltd & Broadcasting & $0.0718$ & $0.0511$ & No\\
$000930$ & Anhui BBCA Biochemical Co., Ltd & Biology & $0.0699$ & $0.0518$ & No \\
$000983$ & Shanxi Xishan Coal and Electricity Power Co., Ltd & Power electronics & $0.0509$ & $0.0496$ & No\\
$600000$ & Shanghai Pudong Development Bank Co., Ltd & Financials & $0.0699$ & $0.0538$ & No\\
$600019$ & Baoshan Iron \& Steel CO., LTD & Metals & $0.0223$ & $0.0672$ &Yes\\
$600026$ & China Shipping Development Co., Ltd & Transportation & $0.0277$ & $0.0511$ &Yes\\
$600028$ & China Petroleum \& Chemical Co., Ltd & Petroleum \& Chemistry & $0.0670$ & $0.0653$ & No\\
$600030$ & CITIC Securities Company Co., Ltd & Financials & $0.0580$ & $0.0520$ & No\\
$600036$ & China Merchants Bank Co., Ltd & Financials & $0.0436$ & $0.0576$ &Yes\\
$600073$ & Shanghai Maling Aquarius Co., Ltd & Food & $0.0501$ & $0.0500$ & No\\
$600088$ & China Television Media Co., Ltd & Broadcasting & $0.0540$ & $0.0497$ & No\\
$600100$ & Tsinghua Tongfang Co., Ltd & IT & $0.0462$ & $0.0500$ &Yes\\
$600104$ & SAIC Motor CO., LTD & Manufacturing & $0.0149$ & $0.0573$ &Yes\\
$600110$ & China-Kinwa High Technology Co., Ltd & Manufacturing & $0.0911$ & $0.0506$ & No\\
$600171$ & Shanghai Belling Co., Ltd & Electronics & $0.0377$ & $0.0530$ &Yes\\
$600320$ & Shanghai Zhenhua Port Machinery Co., Ltd & Manufacturing & $0.0515$ & $0.0506$ & No\\
$600428$ & COSCO Shipping Co., Ltd & Transportation & $0.0475$ & $0.0510$ &Yes\\
$600550$ & Baoding Tianwei Baobian Electric Co., Ltd & Manufacturing & $0.0198$ & $0.0498$ &Yes\\
$600601$ & Founder Technology Group Co., Ltd & IT & $0.0218$ & $0.0637$ &Yes\\
$600602$ & SVA Electron Co., Ltd & Electronics & $0.0167$ & $0.0561$ &Yes\\
$600688$ & SINOPEC Shanghai Petrochemicl Co., Ltd & Petroleum \& Chemistry & $0.0421$ & $0.0566$ &Yes\\
$600770$ & Jiangsu Zongyi Co., Ltd & Conglomerates & $0.0647$ & $0.0478$ & No\\
$600797$ & INSIGMA Technology Co., Ltd & IT & $0.1179$ & $0.0647$ & No\\
$600832$ & Shanghai Oriental Pearl (Group) Co.£¬Ltd & Conglomerates & $0.0610$ & $0.0474$ & No\\
$600900$ & China Yangtze Power Co., Ltd & Power electronics & $0.0574$ & $0.0556$ & No\\\hline %
\end{tabular}
\end{table*}

\begin{figure}[htb]
\centering
\includegraphics[width=8cm]{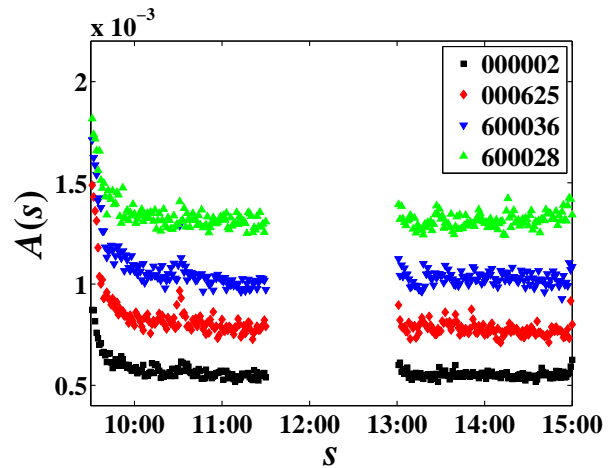}
\caption{(Color online) The 1-min interval intraday patterns of four
representative stocks: $000002$, $000625$, $600036$ and $600028$.
The curves are vertically shifted for clarity.}
\label{Fig:IntradayPattern}
\end{figure}

For most stock markets, the intraday volatilities exhibit a U-shaped
or L-shaped intraday pattern. This is so for the Chinese stocks
\cite{Ni-Zhou-2007-XXX}. When dealing with intraday data, this
pattern should be removed
\cite{Wang-Yamasaki-Havlin-Stanley-2006-PRE,Wang-Weber-Yamasaki-Havlin-Stanley-2007-EPJB,VodenskaChitkushev-Wang-Weber-Yamasaki-Havlin-Stanley-2008-EPJB,Wang-Yamasaki-Havlin-Stanley-2008-PRE,Qiu-Guo-Chen-2008-PA}.
Otherwise, the return intervals distribution will exhibit daily
periodicity for large thresholds. The intraday pattern $A(s)$ is
defined as
\begin{equation}
A(s)=\frac{\sum_{i=1}^{N} R^i(s)}{N}, \label{e10}
\end{equation}
which is the volatility at a specific moment $s$ of the trading day
averaged over all $N$ trading days and $R^i(s)$ is the volatility at
time $s$ of day $i$. The L-shaped intraday patterns of four typical
stocks are illustrated in Figure \ref{Fig:IntradayPattern}. The
intraday pattern exhibits a pronounced peak at the opening hours,
and a minimum value just before the midday break of a trading day.
In comparison with the western stock markets, the intraday pattern
of the Chinese stock market surges soon after midday break which may
due to the information aggregation during the midday break.

To avoid the effect of this daily oscillation, the intraday pattern
is removed by
\begin{equation}
R'(t)=\frac{R(t)}{A(s)}. \label{e20}
\end{equation}
Then we normalize the volatility by dividing its standard deviation
\begin{equation}
r(t)=\frac{R'(t)}{[\langle R'(t)^2 \rangle-\langle R'(t)
\rangle^2]^{1/2}}. \label{e30}
\end{equation}
We note that the raw 1-min Chinese stock returns follow Student
distribution whose tail exponent is about 3.5
\cite{Gu-Chen-Zhou-2008a-PA}, which ensures that the variance of
$R'(t)$ exists where the intraday pattern is removed.

\begin{figure}[htb]
\centering
\includegraphics[width=8cm]{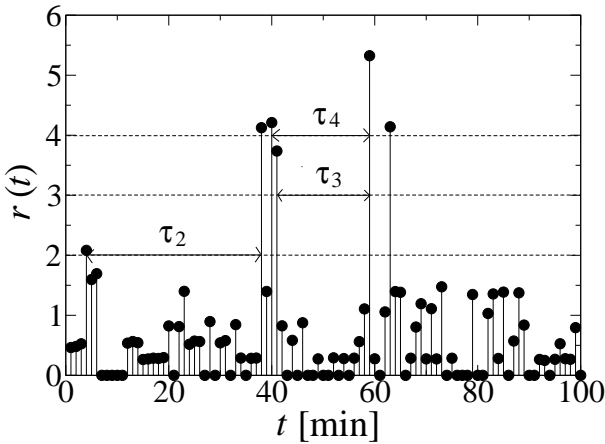}
\caption{Illustration of volatility return intervals for stock
$000625$, where $\tau_2$, $\tau_3$ and $\tau_4$ correspond to return
intervals for $q=2$, $3$ and $4$.} \label{Fig:VRI}
\end{figure}

We study the return intervals $\tau$ between successive volatilities
$|r|$ exceeding a certain threshold $q$. The definition of return
intervals is illustrated in Figure \ref {Fig:VRI} for a typical
stock. It is clear that the number of return intervals corresponding
to a given threshold decreases with increasing threshold and the sum
of return intervals $\tau_q$ are approximately identical to each
other for different $q$. When the threshold $q$ is very small, say
less than the minimum of $r(t)$, then all $\tau_q$ values equal to
1.

\section{Scaling and nonscaling behaviors of return interval distributions}
\label{S1:PDF}

Several empirical studies show that the probability distribution
function (PDF) of the return intervals obeys a scaling form:
\begin{equation}
P_q(\tau)=\frac{1}{\bar{\tau}} f \left( \frac{\tau}{\bar{\tau}}
\right), \label{Eq:Pq:f}
\end{equation}
where $\bar{\tau}$ is the mean return interval which depends on the
threshold $q$. The scaling form could be approximated by a stretched
exponential function as
\begin{equation}
f(x)=c e^{- \alpha x^{\gamma}}, \label{Eq:StrExp}
\end{equation}
where $c$ and $\alpha$ are two parameters and $\gamma$ is the
correlation exponent characterizing the long-term memory of
volatilities. In this section, we investigate if these two
properties hold for the 30 Chinese stocks.

\subsection{Complementary cumulative distributions}
\label{S2:PDF:cdf}

\begin{figure*}[htb]
\centering
\includegraphics[width=8cm]{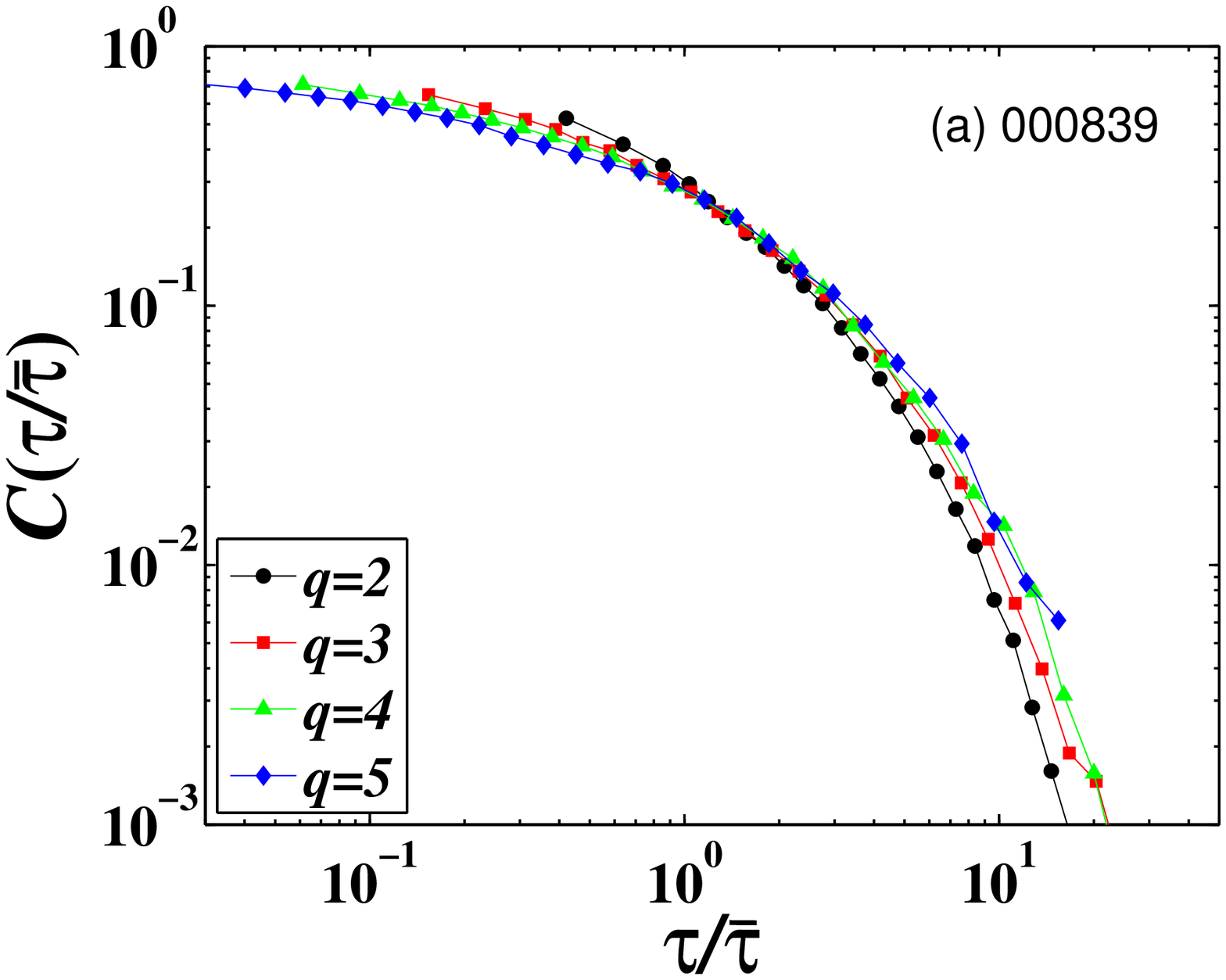}\hspace{5mm}
\includegraphics[width=8cm]{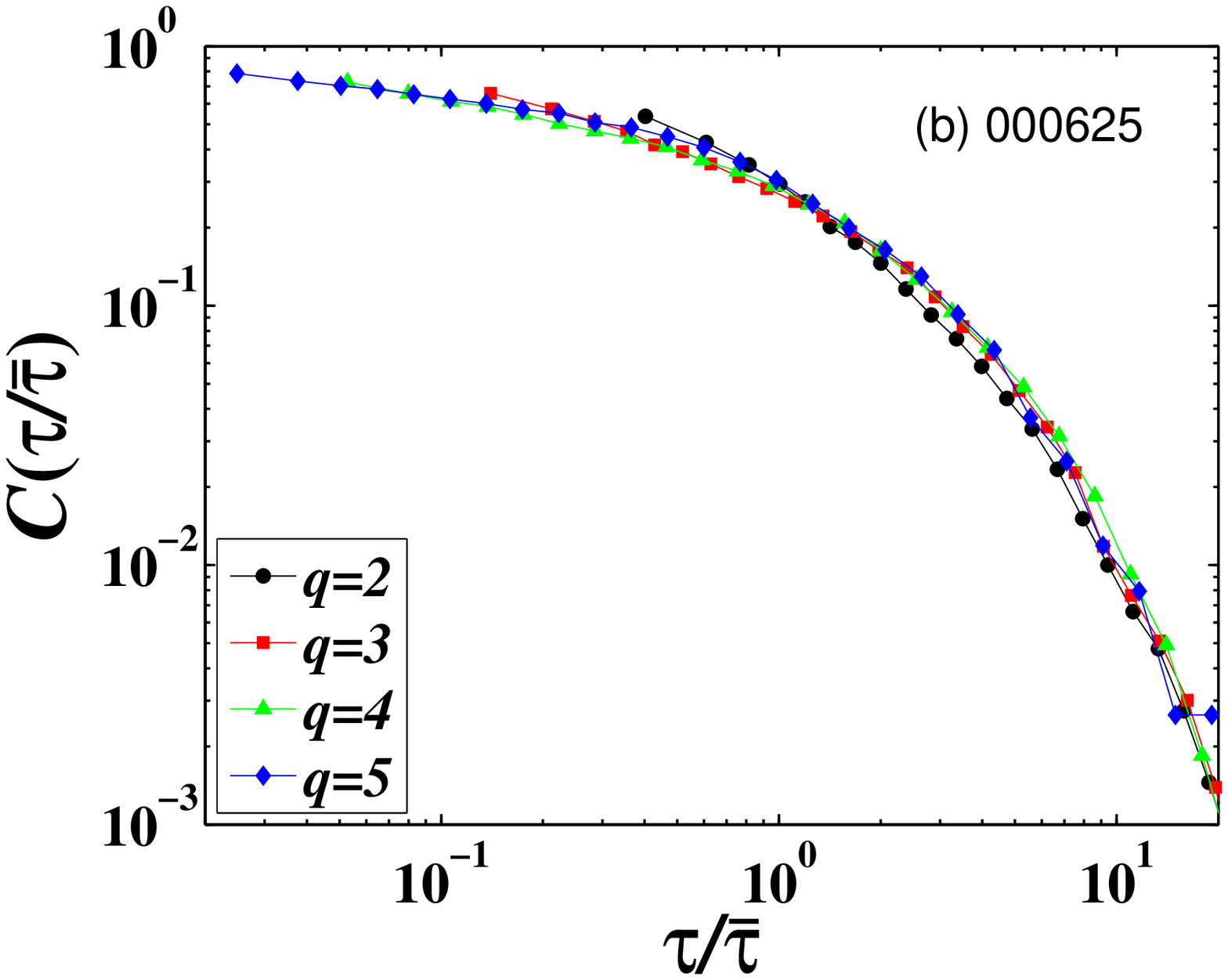}
\caption{(Color online) Empirical complementary cumulative
distribution of scaled return intervals for different thresholds
$q=2,3,4,5$ for (a) stock 000839 which shows clear deviations from
scaling, and (b) stock 000625 which apparently shows scaling
behavior.} \label{Fig:PDF:Scaling}
\end{figure*}

To make the observation of this possible scaling behavior
(\ref{Eq:Pq:f}) more clear, we study the complementary cumulative
distribution function (CCDF) of the scaled return intervals
\begin{equation}
C_q(\tau/\bar{\tau})= \int_\tau^\infty P_q(\tau)d\tau
=\int_{\tau/\bar{\tau}}^\infty f(x)dx. \label{e60}
\end{equation}
If the PDFs for different $q$ obey the scaling form in Eq.
(\ref{Eq:Pq:f}), the CCDFs of the scaled return intervals should
also collapse onto a single curve.

We calculate the CCDFs of the scaled return intervals of the chosen
$30$ stocks for a wide range of threshold $q=2,3,4,5$. Note that $q$
is in units of standard deviations such that $q=5$ is a quite large
volatility. The CCDFs of a representative stock 000839 are shown in
Figure \ref{Fig:PDF:Scaling}(a). The curves for different values of
threshold $q$ seem close to each other but do not collapse onto a
single curve. A systematic deviation from scaling is clearly
observed: the cumulative distribution decreases with the increase of
threshold $q$ for small scaled return intervals $\tau/\bar{\tau}<1$,
while it increases with $q$ for large scaled return intervals
$\tau/\bar{\tau}>1$. This observation is similar to the multiscaling
behavior reported by Wang {\em{et al.}}
\cite{Wang-Yamasaki-Havlin-Stanley-2008-PRE}.

However, we also find some stocks which apparently show scaling
behaviors. In Figure \ref{Fig:PDF:Scaling}(b), the CCDFs of a
representative stock 000625 that exhibits approximate scaling
behavior are plotted. One observes that the curves for different
values of threshold $q$ approximately collapse onto a single curve.
This indicates the return interval distributions of this stock obey
a scaling behavior.

\subsection{Kolmogorov-Smirnov test of scaling in return interval distributions}
\label{S2:PDF:KS:Scaling}

The eyeballing of complementary cumulative distributions offers a
qualitative way of distinguishing scaling and nonscaling behaviors.
Here we further adopt a quantitative approach, the
Kolmogorov-Smirnov (KS) test. We use the KS test to compare two
distributions for $q=2$ and $q=5$, which behave most differently
among all the $q$ value distributions. If the distributions for
$q=2$ and $q=5$ pass the test, we can conclude that all the return
interval distributions for different $q$ values collapse onto a
single curve and thus obey a scaling law.

The standard KS test is designed to test the hypothesis that the
distribution of the empirical data is equal to a particular
distribution by comparing their cumulative distribution functions
(CDFs). Our hypothesis is that the two return interval distributions
for $q=2$ and $q=5$ do not differ at least in the common region of
the scaled return intervals
\cite{Jung-Wang-Havlin-Kaizoji-Moon-Stanley-2008-EPJB}. Suppose that
$F_2=1-C_2$ is the CDF of return intervals for $q=2$ and $F_5=1-C_5$
is the CDF of return intervals for $q=5$. We calculate the KS
statistic by comparing the two CDFs in the overlapping region:
\begin{equation}
   KS = \max\left(|F_2-F_5|\right)~.
   \label{Eq:KS}
\end{equation}
When the KS statistic is smaller than a critical value denoted by
$CV$ (i.e., $KS<CV$), the hypothesis is accepted and we can assume
that the distribution for $q=2$ is coincident with the distribution
for $q=5$. The critical value at the significance level of 5\% is
$CV=1.36/\sqrt{{mn}/({m+n})}$, where $m$ and $n$ are the numbers of
interval samples for $q=2$ and $5$, respectively. In
Table~\ref{TB:two-sample-KS} is depicted the KS statistics and the
corresponding critical values for the $30$ stocks. One can see that
$12$ stocks including the stock shown in
Figure~\ref{Fig:PDF:Scaling}(b) pass the KS test and consequently
their return interval distributions obey scaling behaviors, while
other 18 stocks including the one shown in
Figure~\ref{Fig:PDF:Scaling}(a) do not pass the test. We find that
stock 600030, which is the only stock also studied by Qiu {\em{et
al.}} \cite{Qiu-Guo-Chen-2008-PA}, does not pass the KS test at the
significance level of 5\%, but the statistic $KS$ is very close to
the critical value $CV$, i.e., $0.0580\sim0.0520$. Further study
shows that the stock 600030 passes the KS test at a significance
level of $1.97\%$ which indicates this stock may follow a relatively
week scaling behavior.

\subsection{Kolmogorov-Smirnov test of the scaling function}
\label{S2:PDF:KS:Fun}

We have demonstrated that the return distributions of $12$ stocks
show scaling behaviors. To further study the particular form of the
scaling function, we perform the KS goodness-of-fit test
\cite{Clauset-Shalizi-Newman-2007-XXX,Gonzalez-Hidalgo-Barabasi-2008-Nature}.
Empirical studies have shown that the scaling form could be
approximated by a stretched exponential function as in
Eq.~(\ref{Eq:StrExp}). In this case, our hypothesis is that the
empirical distribution is coincident with the fitted stretched
exponential distribution. Similar to the KS test we have conducted
for two empirical samples, we use the KS statistics to test whether
the distributions for $q=2$ and $q=5$, which behave most differently
among all the $q$ value distributions, are identical to a same
fitted distribution for both $q$ values in the overlapping region of
the scaled return intervals. If both return interval distributions
for $q=2$ and $q=5$ pass the test, we can conclude that all the
return interval distributions for different $q$ values follow a
scaling function with stretched exponential form.

The method is described as follows. Let $F_q=1-C_q$ be the
cumulative distribution of $\tau_q$, $F_{SE}$ the cumulative
distribution from the fitted stretched exponential. The KS statistic
defined in Eq.~(\ref{Eq:KS}) becomes
\begin{equation}
   KS = \max\left(|F_q-F_{\rm{SE}}|\right),~~~~q=2,5~.
   \label{Eq:KS2}
\end{equation}
We also use an variant of the KS statistic known as the weighted KS
statistic, which is defined as follows
\cite{Gonzalez-Hidalgo-Barabasi-2008-Nature}
\begin{equation}
KSW = \max
\left(\frac{|F_q-F_{\rm{SE}}|}{\sqrt{F_{\rm{SE}}(1-F_{\rm{SE}})}}
\right), \label{Eq:KSw}
\end{equation}
which is more sensitive on the edges of the cumulative distribution.
In this case, the bootstrapping approach is adopted
\cite{Clauset-Shalizi-Newman-2007-XXX,Gonzalez-Hidalgo-Barabasi-2008-Nature}.
To do this, we first generate 1000 synthetic samples from the best
fitted distribution and then reconstruct the cumulative distribution
$F_{\rm{sim}}$ of each simulated sample and its CDF
$F_{\rm{sim,SE}}$ from integrating the fitted stretched exponential.
We calculate the values of $KS$ and $KSW$ between the fitted CDF and
the simulated CDF using
\begin{equation}
   KS_{\rm{sim}} = \max\left(|F_{\rm{sim}}-F_{\rm{sim,SE}}|\right)
   \label{Eq:KS2:sim}
\end{equation}
and
\begin{equation}
 KSW_{\rm{sim}} = \max
 \left(\frac{|F_{\rm{sim}}-F_{\rm{sim,SE}}|}{\sqrt{F_{\rm{sim,SE}}(1-F_{\rm{sim,SE}})}}\right).
 \label{Eq:KSw:sim}
\end{equation}
The $p$-value is determined by the frequency that $KS_{\rm{sim}}>KS$
or $KSW_{\rm{sim}}>KSW$. The tests are carried out for the 12 stocks
exhibiting scaling behaviors. The resultant $p$-values are depicted
in Table \ref{TB:goodness-of-fit-KS}.

\begin{table}[htp]
 \centering
 \caption{$KS$ and $KSW$ test of the return interval
distributions for $q=2$ and $5$ by comparing empirical data with the
best fitted distribution and synthetic data with the best fitted
distribution. } \label{TB:goodness-of-fit-KS}
\begin{tabular}{cccccccc}
  \hline
   Code & $q$ & $p_{KS}$ & $p_{KSW}$ & Code & $q$ & $p_{KS}$ & $p_{KSW}$\\
  \hline
  $000625$ & $2$ & $0.224$ & $0.769$ & $600019$ & $2$ & $0.938$ & $0.742$\\\vspace{1mm}%
           & $5$ & $0.045$ & $0.023$ &          & $5$ & $0.567$ & $0.574$\\
  $600026$ & $2$ & $0.010$ & $0.215$ & $600036$ & $2$ & $0.311$ & $0.432$\\\vspace{1mm}%
           & $5$ & $0.021$ & $0.011$ &          & $5$ & $0.086$ & $0.098$\\
  $600100$ & $2$ & $0.003$ & $0.022$ & $600104$ & $2$ & $0.851$ & $0.973$\\\vspace{1mm}%
           & $5$ & $0.157$ & $0.184$ &          & $5$ & $0.709$ & $0.866$\\
  $600171$ & $2$ & $0.158$ & $0.286$ & $600428$ & $2$ & $0.004$ & $0.001$\\\vspace{1mm}%
           & $5$ & $0.401$ & $0.346$ &          & $5$ & $0.002$ & $0.007$\\
  $600550$ & $2$ & $0.226$ & $0.354$ & $600601$ & $2$ & $0$     & $0$\\\vspace{1mm}%
           & $5$ & $0.059$ & $0.103$ &          & $5$ & $0.883$ & $0.130$\\
  $600602$ & $2$ & $0.995$ & $0.943$ & $600688$ & $2$ & $0.038$ & $0.386$\\
           & $5$ & $0.996$ & $0.592$ &          & $5$ & $0.188$ & $0.056$\\
  \hline
\end{tabular}
\end{table}

Consider the significance level of 1\%. If at least one $p$-value
for $q=2$ or $q=5$ of an individual stock is less than 1\%, then the
null hypothesis that the empirical PDFs of this stock can be well
fitted by a stretched exponential is rejected. According to
Table~\ref{TB:goodness-of-fit-KS}, the null hypothesis is rejected
for three stocks (600100, 600428, 600601) using the KS test and for
two stocks (600428, 600601) using the KSW test. Under the
significant level of 5\%, the null hypothesis is rejected for six
stocks (000625, 600026, 600100, 600428, 600601, 600688) using the KS
test and for five stocks (000625, 600026, 600100, 600428, 600601)
using the KSW test. We find that the KS test and the KSW test
provide very similar results except that the KS test is slightly
stronger \cite{Gonzalez-Hidalgo-Barabasi-2008-Nature}. It is
noteworthy to point out that the $p$-values of stocks 600602, 600019
and 600104 are very large, implying high goodness-of-fit of the
stretched exponential to the empirical PDFs. In principle, the
$p$-values of a stock are larger when the scaling of PDFs for
different $q$ is more significant, which is manifested by the fact
that the statistic $KS$ exceeds the critical value $CV$ more in
Table \ref{TB:two-sample-KS}.

\begin{figure}[htb]
\centering
\includegraphics[width=8cm]{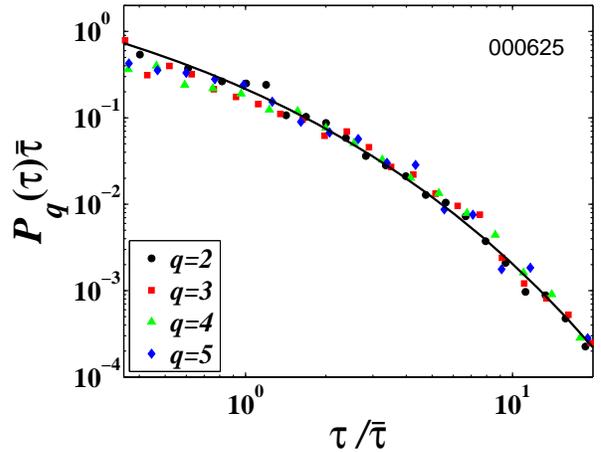}
\caption{(Color online) Scaled probability distribution of scaled
return intervals for different threshold $q=2,3,4,5$ for the
representative stock 000625, which show scaling behavior. The solid
curve is the fitted function $c e^{- \alpha x^{\gamma}}$.}
\label{Fig:Fit:SE}
\end{figure}

In all the fitting, the correlation exponent $\gamma$ for different
stocks lies in a range $[0.22,0.44]$ with an average value $0.31$,
which is consistent with that observed in
\cite{Qiu-Zheng-Ren-Trimper-2007-PA}. To show how good the stretched
exponential fits the data, we illustrate in Figure~\ref{Fig:Fit:SE}
the empirical PDFs of $\tau_q$ and the fitted curve for a
representative stock $000625$, which passes the KS and KSW
goodness-of-fit tests at the significance level of 1\% but falls at
the significance level of 5\%. It is obviously that the curves for
different values of $q$ almost collapse onto a single curve which
could be nicely fitted by a stretched exponential function.

\section{Memory effect of return intervals}
\label{S1:Memory}

The probability distribution may not fully characterize the
properties of volatility return intervals. The temporal correlation
is known as another important observable independent of the
probability distribution. Empirical studies have revealed that the
stock market volatilities are long-term correlated. We suppose that
the long-term correlated volatilities would also affect the memory
of the return intervals. Indeed, the memory effect is observed in
the return intervals of US stock market and Japanese stock market.

\subsection{Short-term memory of return intervals}
\label{S2:Memory:Short}

\begin{figure}[htb]
\centering
\includegraphics[width=8cm]{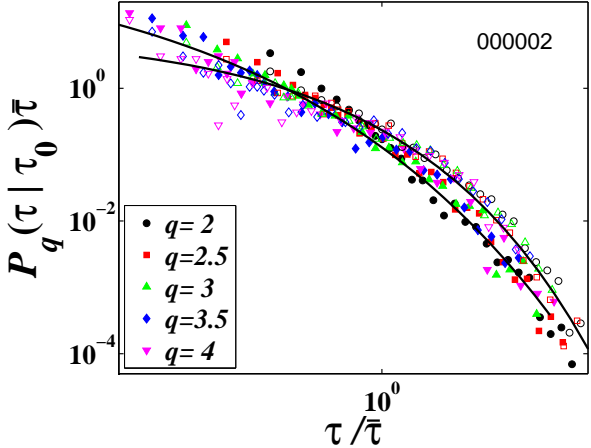}
\caption{(Color online) Scaled conditional PDF $P_q(\tau | \tau_0)
\bar{\tau}$ of scaled return intervals $\tau/\bar{\tau}$ with
$\tau_0$ in the largest $1/4$ subset (open symbols) and smallest
$1/4$ subset (filled symbols) for a representative stock 000002. The
solid lines are fitted curves.} \label{Fig:Pr:Cond}
\end{figure}

To investigate the memory effect of the return intervals in Chinese
stock market, we first calculate the conditional probability
distribution $P_q(\tau|\tau_0)$, which is the probability to find an
interval $\tau$ immediately after the interval $\tau_0$.
Specifically, we study the conditional PDF for a bin of $\tau_0$.
The entire interval sequences are arranged in ascending order and
partitioned to four bins with equal size. We illustrate the results
using a typical stock 000002. The results for all other 29 stocks
are qualitatively the same. Figure~\ref{Fig:Pr:Cond} plots
$P_q(\tau|\tau_0)$ as a function of the scaled return intervals
$\tau/\bar{\tau}$ for $\tau_0$ in the smallest and biggest bins. For
large $\tau/\bar{\tau}$, $P_q(\tau|\tau_0)$ with $\tau_0$ in the
biggest subset is larger than $P_q(\tau|\tau_0)$ with $\tau_0$ in
the smallest subset, and for small $\tau/\bar{\tau}$,
$P_q(\tau|\tau_0)$ with $\tau_0$ in the biggest subset is smaller
than $P_q(\tau|\tau_0)$ with $\tau_0$ in the smallest subset. This
indicates that small intervals $\tau_0$ tend to be followed by small
intervals $\tau$, while large intervals $\tau_0$ tend to be followed
by large intervals $\tau$. Therefore, there exists short-term memory
in the return intervals.

\begin{figure}[htb]
\centering
\includegraphics[width=8cm]{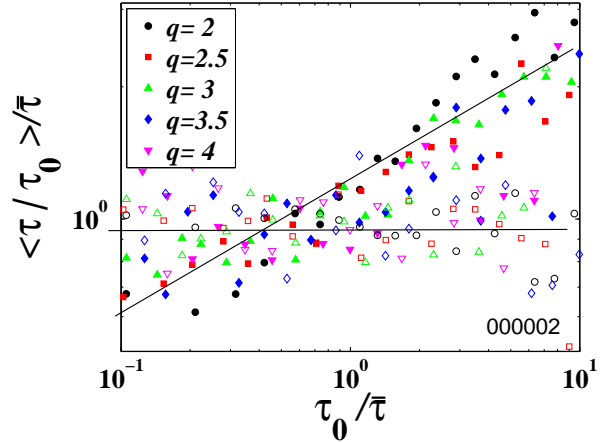}
\caption{(Color online) Scaled mean conditional return interval
$\langle \tau| \tau_0 \rangle/\bar{\tau}$ of scaled return intervals
(filled symbols) and shuffled data (open symbols) for a typical
stock 000002. The solid lines are guidelines.} \label{Fig:mean:tau}
\end{figure}

Calculating the mean conditional return interval $\langle \tau|
\tau_0 \rangle$ immediately after $\tau_0$ is another alternative
way to investigate the memory of return intervals. In fact, $\langle
\tau| \tau_0 \rangle$ is the first moment of $P_q(\tau | \tau_0)$.
In Figure \ref{Fig:mean:tau} is shown $\langle \tau| \tau_0 \rangle$
of stock 000002, which exhibits a monotonously increasing tendency
with the increase of the scaled return interval $\tau/\bar{\tau}$.
This also indicates that small $\tau$ tends to follow small $\tau_0$
and big $\tau$ tends to follow big $\tau_0$. This result is
consistent with that observed in $P_q(\tau|\tau_0)$. For those
shuffled return intervals, $\langle \tau| \tau_0 \rangle$ fluctuates
around a horizontal line close to $1$ indicating that $\tau$ is
independent of the previous $\tau_0$, as expected. This suggests
that the memory of the return intervals may arise from the long-term
memory of the volatilities.

\subsection{Long-term memory of return intervals}
\label{S2:Memory:Long}

\begin{figure}[htb]
\centering
\includegraphics[width=8cm]{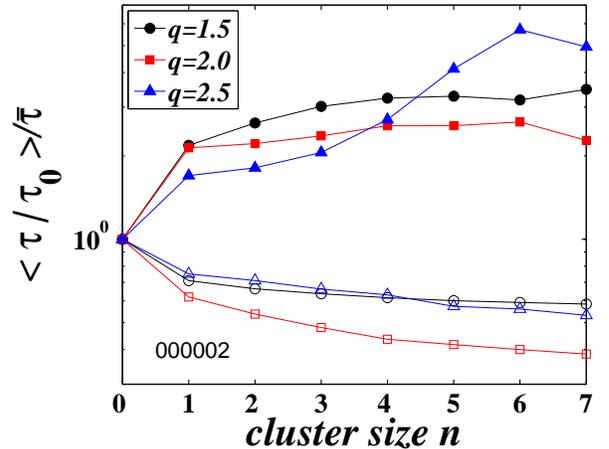}
\caption{(Color online) Scaled mean return interval $\langle \tau|
\tau_0 \rangle/\bar{\tau}$ after ``+'' cluster (filled symbols) and
``-'' cluster (open symbols) for a representative stock 000002.}
\label{Fig:Cluster}
\end{figure}

\begin{figure*}[htb]
\centering
\includegraphics[width=7cm,height=6cm]{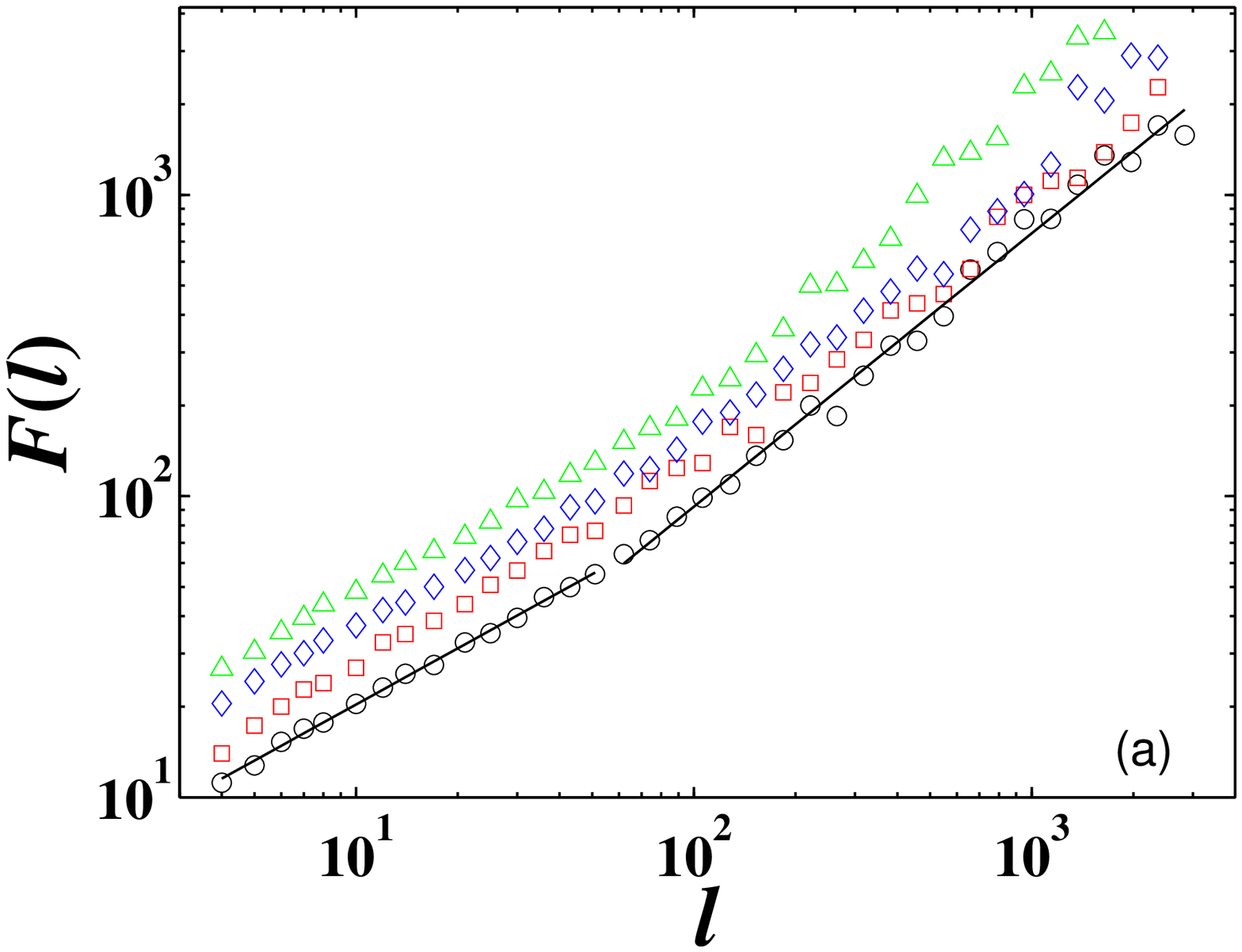}\hspace{5mm}
\includegraphics[width=7cm,height=6cm]{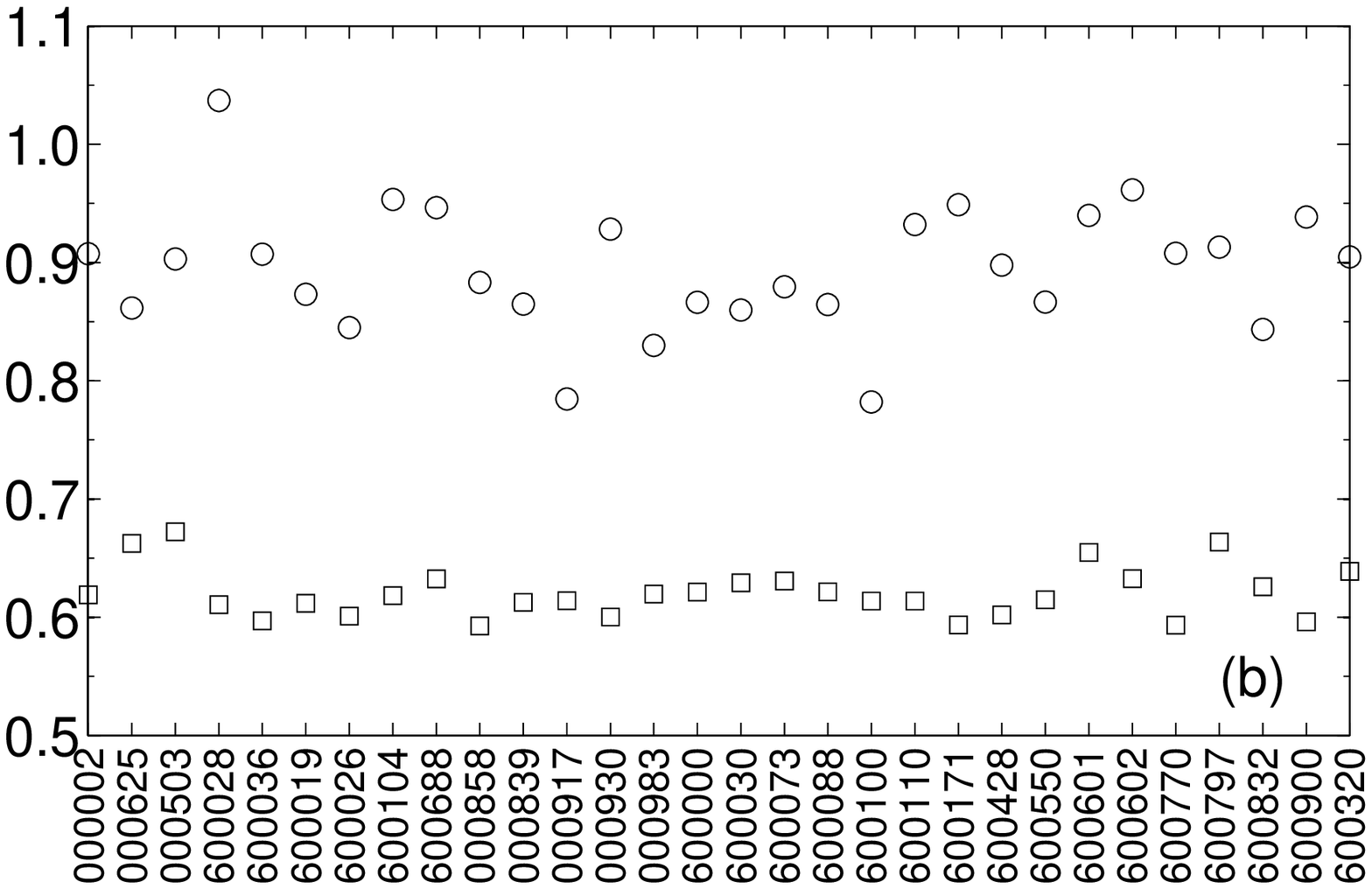}
\caption{\label{Fig:DFA} (a) Detrended fluctuation $F(l)$ of return
interval $\tau_2$ for four representative stocks. The curves are
vertically shifted for clarity. (b) Extracted Hurst exponents $H$
for large scales (circles) and small scales (squares) for the $30$
stocks.}
\end{figure*}

To further study the long-term memory of return intervals, we
investigate the mean return interval $\langle \tau| \tau_0 \rangle$
after a cluster of $n$ intervals that are all in a bin $\tau_0$. The
entire interval sequences are partitioned into two bins, separated
by the median value of return intervals. Thus we investigate the
mean return interval after a cluster of $n$ intervals consecutively
above and below the median value, denoted by ``+'' and ``-''
clusters. Figure \ref{Fig:Cluster} illustrates $\langle \tau| \tau_0
\rangle$ with respect to the cluster size $n$ for a representative
stock 000002. The results for other stocks are similar. In general,
$\langle \tau| \tau_0 \rangle$ after ``+'' cluster increases, while
$\langle \tau| \tau_0 \rangle$ after ``-'' cluster decreases as the
cluster size $n$ increases. In other words, a large mean return
interval is likely followed by large consecutive return intervals
and a small return interval is likely followed by small consecutive
return intervals. This observation shows the dependence of the
return interval on its previous $n$ consecutive return intervals.

The detrended fluctuation analysis (DFA) method is applied to
directly investigate the long-term correlation of return intervals.
After removing a linear trend, the DFA method computes the detrended
fluctuation $F(l)$ of the time series within a window of $l$ points
and determines the exponent $H$ from the scaling form $F(l)\sim l^
H$, where $H$ is the Hurst exponent
\cite{Peng-Buldyrev-Havlin-Simons-Stanley-Goldberger-1994-PRE,Kantelhardt-Bunde-Rego-Havlin-Bunde-2001-PA}.
For $H>0.5$ the time series is long-term correlated, and for $H=0.5$
the time series is uncorrelated. The detrended fluctuation function
$F(l)$ of four representative stocks are shown in
Figure~\ref{Fig:DFA}(a). A crossover behavior is observed: $F(l)$
for small scales $l<50$ obeys a power law with an relatively small
exponent and $F(l)$ for large scales $l>50$ obeys a power law with a
relatively large exponent. This phenomenon is also observed for
other stocks, which is in line with other stock markets
\cite{Wang-Yamasaki-Havlin-Stanley-2006-PRE}.
Figure~\ref{Fig:DFA}(b) illustrates the estimated Hurst exponents
$H$ of the $30$ stocks. In both regions for small scales and large
scales, the Hurst exponents of all the individual stocks are
apparently greater than $0.5$, which indicates the existence of
long-term memory in return intervals. For shuffled data, the Hurst
exponent displays a value close to $0.5$. This confirms that the
long-term memory may arise from the long-term memory of volatility
records.

\section{Summary and conclusions}
\label{S1:concl}

In summary, we have studied the distribution and memory effect of
volatility return intervals for 30 most actively traded stocks on
the Shanghai and Shenzhen Stock Exchanges. The Kolmogorov-Smirnov
tests are performed to examine the scaling behavior of the return
interval distributions as well as the particular form of the scaling
function. We find that 12 stocks exhibit scaling behaviors in the
volatility return interval distributions. We further find that the
scaling form of 6 stocks having the best scaling can be excellently
approximated by a stretched exponential function $f(x) \sim e^{-
\alpha x^{\gamma}}$, and the mean value of the correlation exponent
$\gamma$ is estimated to be around $0.31$.

The memory effect is also observed in the volatility return
intervals of the $30$ stocks. We first calculate the conditional PDF
$P_q(\tau | \tau_0)$ and mean conditional return interval $\langle
\tau| \tau_0 \rangle$, and find that the interval $\tau$ is
dependent of its closest previous interval $\tau_0$, which indicates
the short-term memory of return intervals. We further investigate
$\langle \tau| \tau_0 \rangle$ after a cluster of $n$ intervals
consecutively above and below the median value, and find that memory
exists in a cluster of long return intervals. In addition, detrended
fluctuation analysis of the return intervals shows that the return
intervals have long-term memory.

In a nutshell, our work focuses on the study of the statistical
properties of volatility return intervals of Chinese stocks. Our
main finding is that some stocks show scaling behaviors while the
others do not. This novel feature shows the complexity of the
Chinese stock market and help us better understand the properties of
volatility return intervals.

\begin{acknowledgement}
We thank Z.-Q. Jiang, G.-F. Gu, G.-H. Mu and T. Qiu for helpful
discussions and suggestions. This work was partially supported by
the Shanghai Educational Development Foundation (No. 2008CG37), the
National Natural Science Foundation of China (No. 70501011), the Fok
Ying Tong Education Foundation (No. 101086), and the Program for New
Century Excellent Talents in University (No. NCET-07-0288).
\end{acknowledgement}

\bibliographystyle{epj}
\bibliography{E:/papers/Auxiliary/Bibliography}

\end{document}